\documentclass{jfm}

\usepackage{graphicx}
\usepackage{amssymb, amsmath}
\usepackage{subfigure}
\usepackage{epstopdf}
\usepackage{natbib}

\usepackage{color}
\usepackage[colorinlistoftodos]{todonotes}
\ifCUPmtlplainloaded \else
  \checkfont{eurm10}
  \iffontfound
    \IfFileExists{upmath.cls}
      {\typeout{^^JFound AMS Euler Roman fonts on the system,
                   using the 'upmath' package.^^J}%
       \usepackage{upmath}}
      {\typeout{^^JFound AMS Euler Roman fonts on the system, but you
                   dont seem to have the}%
       \typeout{'upmath' package installed. JFM.cls can take advantage
                 of these fonts,^^Jif you use 'upmath' package.^^J}%
      }
  \else
  \fi
\fi

\ifCUPmtlplainloaded \else
  \checkfont{msam10}
  \iffontfound
    \IfFileExists{amssymb.sty}
      {\typeout{^^JFound AMS Symbol fonts on the system, using the
                'amssymb' package.^^J}%
       \usepackage{amssymb}%
       \let\le=\leqslant  \let\leq=\leqslant
         
      }{}
  \fi
\fi

\ifCUPmtlplainloaded \else
  \IfFileExists{amsbsy.sty}
    {\typeout{^^JFound the 'amsbsy' package on the system, using it.^^J}%
     \usepackage{amsbsy}}
    {\providecommand\boldsymbol[1]{\mbox{\boldmath $##1$}}}
\fi

\newsavebox{\astrutbox}
\sbox{\astrutbox}{\rule[-5pt]{0pt}{20pt}}

\title[Predicting the breaking strength of gravity water waves]{Predicting the breaking strength of gravity water waves}
\author[M. Derakhti,   M. L. Banner, and J. T. Kirby]%
{M\ls O\ls R\ls T\ls E\ls Z\ls A\ns D\ls E\ls R\ls A\ls K\ls H\ls T\ls I\ls$^1$
  \thanks{Email address for correspondence: derakhti@jhu.edu}
  M\ls I\ls C\ls H\ls A\ls E\ls L\ns L.\ns B\ls A\ls N\ls N\ls E\ls R\ls$^2$\;\and\;
  \; J\ls A\ls M\ls E\ls S\ns T.\ns  K\ls I\ls R\ls B\ls Y$^3$ 
}
\affiliation{$^1$Department of Civil Engineering, Johns Hopkins University, Baltimore, MD 21218, USA \\
$^2$School of Mathematics and Statistics, The University of New South Wales Sydney, Sydney 2052 Australia \\
$^3$ Center for Applied Coastal Research, Department of Civil and Environmental Engineerings, University of Delaware, Newark, DE 19716, USA  \\
[\affilskip]}
\pubyear{?}
\volume{?}
\pagerange{?--?}
\date{?; revised ?; accepted ?.}

\begin{document}

\maketitle

\begin{abstract}
We revisit the classical but as yet unresolved problem of predicting the strength of breaking  2-D and 3-D gravity water waves, as quantified by the amount of wave energy dissipated per breaking event. Following  \citet{Duncan:1983}, the wave energy dissipation rate per unit length of breaking crest may be related to the fifth moment of the wave speed and the non-dimensional breaking strength parameter $b$.  We use a finite-volume Navier-Stokes solver with LES resolution and volume-of-fluid surface reconstruction \citep{Derakhti-Kirby:2014, derakhti-kirby-jfm16} to simulate wave packet evolution, breaking onset and post-breaking evolution for representative cases of wave packets with breaking due to dispersive focusing and to modulational instability. The present study uses these results to investigate the relationship between the breaking strength parameter $b$ and the breaking onset parameter $B$ proposed recently by \citet{Barthelemy-etal:2018}.  The latter, formed from the local energy flux normalized by the local energy density and the local crest speed, simplifies, on the wave surface, to the ratio of fluid speed to crest speed. Following a wave crest, when $B$ exceeds a generic threshold value at the wave crest \citep{Barthelemy-etal:2018}, breaking is imminent. We find a robust relationship between the breaking strength parameter $b$ and the rate of change of breaking onset parameter, $dB/dt$ at the wave crest, as it transitions through the generic breaking onset threshold ($B\sim 0.85$), scaled by the local period of the breaking wave. This result significantly refines previous efforts to express $b$ in terms of a wave packet steepness parameter, which is both difficult to define robustly and which does not provide a generically accurate forecast of the energy dissipated by breaking.  
\end{abstract}
\section{Introduction}\label{sec:1}
We revisit the classical but as yet unresolved problem of predicting the breaking strength of 2-D and 3-D gravity water waves. The most commonly used approach is the \citet{Phillips:1985} spectral framework for the breaking crest length per unit area, combined with the \citet{Duncan:1983} scaling argument for the wave energy dissipation rate per unit length of breaking crest, $\epsilon$, given by  
\begin{equation}
\epsilon= b\rho g^{-1} c_b^5.   \label{eq1.1}
\end{equation} 

Here, $\rho$ is the density of the liquid phase, $g$ is the gravitational acceleration, $c_b$ is the phase speed of the breaking wave, and $b$ is the dimensionless constant of proportionality hereafter referred to as the breaking strength parameter. 

Field and laboratory data have shown a strong dependence of $b$, within the scatter of data, on the global wave steepness $S$ calculated on the basis of spectral information for the wave packet \citep{Romero-etal:2012}. \citet{derakhti-kirby-jfm16} provided a new scaling for $b$ based on the spectrally-averaged global steepness of a wave field, where the scatter of data in the new formulation was considerably decreased compared with the existing formulations \citep{Romero-etal:2012}. The formulations of \citet{Romero-etal:2012} and \citet{derakhti-kirby-jfm16} provide estimates of $b>0$ for packets with $S>\sim 0.08$. However, $S$ for non-breaking packets can reach values of 0.3 or higher. Thus, another criterion, based on additional information beyond that provided by parameter $S$, is needed to distinguish between breaking and non-breaking packets. In addition, the existing formulations overpredict $b$ for the marginal breaking focused packets. Another inconsistency in the parameterizations of form $b = F(S)$ is a gap in the $b$ values associated with breaking events due to dispersive focusing and modulational instability \citep[see for example][Figure 1]{Romero-etal:2012}, which results partially from an inconsistent choice of averaging time in the analysis of data from wave packet ($\sim$ the local carrier wave period) and modulated wave train ($\sim$ the wave group period) experiments. 
Finally, although such a parameterization is adaptable via the spectral saturation for use in phase-averaged spectral wind-wave models, the evaluation of $S$ is problematic in phase-resolved models. 

\citet{Barthelemy-etal:2018} showed that highest non-breaking waves were clearly separated from marginally breaking waves by their normalized energy fluxes localized near the crest tip region, and that initial breaking instability occurs within a very compact region centered on the wave crest. On the surface, this reduces to the local ratio of the energy flux velocity (here the fluid velocity) to the crest point velocity for the tallest wave in the evolving group. This provides a robust threshold parameter for breaking onset for 2-D wave packets propagating in uniform, deep or intermediate water depths. Further targeted study of representative cases of the most severe laterally-focused 3-D wave packets in deep and intermediate depth water shows that the threshold remains robust. These numerical findings for 2-D and 3-D cases were closely supported by the  observations of \citet{Saket-etal:2017,Saket-etal:2018}. 

Our goal here is to find a robust and local parameterization to predict the breaking strength of 2-D and 3-D gravity water waves. We use a large-eddy-simulation (LES)/volume-of-fluid (VOF) model \citep{Derakhti-Kirby:2014, derakhti-kirby-jfm16} to simulate wave packet evolution, breaking onset and post-breaking evolution for representative cases of wave packets with breaking due to dispersive focusing and to modulational instability. 
Using these numerical results, we investigate the relationship between the breaking strength parameter $b$ and the breaking onset parameter $B$ proposed by \citet{Barthelemy-etal:2018}.
In this paper, we concentrate on focusing-induced breaking events in wave groups over flat bottom topography and conditions ranging from deep to intermediate depth (depth to wavelength ratio from 0.68 to 0.13). Examination of depth-limited breaking or breaking due to strong opposing currents is left for future study. 

\vspace{-4mm}
\section{Numerical experiments}\label{sec:2}
A detailed description of the polydisperse two-fluid model and boundary conditions used may be found in \citet[][\S2]{Derakhti-Kirby:2014}. Demonstrations of model convergence and performance may be found in \citet{Derakhti-Kirby:2014b,derakhti-kirby-jfm16}. The model parameters for a polydisperse bubble phase are chosen as summarized in \citet[][table 4]{Derakhti-Kirby:2014}. Here, the incident wave boundary condition and model set-up are discussed briefly. 

\begin{table}
  \begin{center}
\def~{\hphantom{0}}
  \begin{tabular}{lccccccccccc}
     Case&	$S$	&$f_c$ 		&${\Delta f }/{f_c}$&$N$ &$h$		&$ h/L_0$	&$\tan\theta_y$	&$c_b$		&$T_b$	& $\Gamma$ & $b$	\\[3pt]
     no.   &		&(s$^{-1}$)	&				&	 &(m)		&		&		&(m/s)		&(s)	&						   			 &	$\times10^{3}$\\[3pt]
     A1    &0.30	&0.88		&0.75          	&32	 & 0.60  	&0.25  	&0				&-			&-		&-					 	& -   \\
     A2    &0.3005	&0.88		&0.75           &32	 & 0.60  	&0.25  	&0				&-			&-		&	-					&  -  \\
     A3    &0.302	&0.88		&0.75           &32	 & 0.60  	&0.25  	&0				&1.71		&1.14	&0.45					&  0.4  \\
     A4    &0.31	&0.88		&0.75           &32	 & 0.60  	&0.25  	&0				&1.64		&1.08	&0.83					&  4.5  \\
     A5    &0.41	&0.88		&0.75           &32	 & 0.60  	&0.25  	&0				&1.59		&1.04	&1.22					&  30.1  \\
     A6    &0.44	&0.88		&0.75           &32	 & 0.60  	&0.25  	&0				&1.59		&1.04	&1.44					&  40.9  \\
     B1    &0.25	&0.88		&0.75           &32	 & 0.25  	&0.13  	&[-0.8,0.8]		&1.35 		&1.30	&1.29					&45.1  \\
     B2    &0.25	&0.88		&0.75           &32	 & 0.60  	&0.25  	&[-0.8,0.8]		&1.76 		&1.16	&1.19					&23.4  \\
     B3    &0.35	&0.88		&0.75           &32	 & 0.60  	&0.25  	&[-0.5,0.5]		&1.68 		&1.06	&1.62					&72.7  \\
     B4    &0.35	&0.88		&0.75           &32	 & 0.85  	&0.30  	&[-0.5,0.5]		&1.80		&1.09	&1.39					&35.3  \\
     C1    &0.32	&0.75		&1.0            &32	 & 0.60 	&0.17  	&0				&1.94		&1.41	&0.65					&  1.6  \\
     C2    &0.36	&0.75		&1.0            &32	 & 0.60  	&0.17  	&	0			&1.91		&1.37	&0.86					&  12.4  \\
     C3    &0.40	&0.75		&1.0            &32	 & 0.60  	&0.17  	&	0			&1.95		&1.43	&1.14					&  22.3  \\
     M1	   &0.160	&1.48		&0.0954			&2	 &0.55		&0.68	&	0			&0.88-0.92	&0.56-0.59 	&0.61-1.07			& 1.4-16.9	\\
     M2	   &0.176	&1.48		&0.0954			&2	 &0.55		&0.68	&	0			&0.89-0.95	&0.57-0.61	&0.46-1.12			& 0.3-17.3 \\
     \end{tabular}
\caption{Input parameters for the simulated alongcrest-uniform (2-D) and short-crested (3-D) focused wave packets and 2-D modulated wave trains. }
  \label{tab1}
  \end{center}
\end{table}

We define the coordinate system $(x,y,z)$ such that $x$ and $y$ represent the along-tank and transverse directions respectively and $z$ is the vertical direction, positive upward and measured from the still water level. The reference time $t^*$ and $x$-location $x^*$ are taken as the time and location at which $B$ following the crest tip reaches the threshold value of 0.85 for breaking packets, or its maximum for non-breaking packets, respectively and are normalized by the local period and wave length of the carrier wave respectively. 

All model simulations are performed with the model initialized with quiescent conditions. An incident wave packet is then generated at the model upstream boundary. The input wave packet was composed of $N$ sinusoidal components of steepness $a_nk_n, n=1,\cdots,N$ where $a_n$ and $k_n$ are the amplitude and wave number of the $n$th frequency component. Based on linear theory, the free surface elevation for the 2-D \citep{Rapp-Melville:1990,Derakhti-Kirby:2014} and 3-D \citep{Wu-Nepf:2002,Kirby-Derakhti:2017} focused packets at the wavemaker is given by \vspace{-1mm}
\begin{equation}
\eta (0,y,t) = \sum_{n=1}^{N}a_n\cos[2\pi f_n(t-t_f)+\frac{k_nx_f}{\cos{\theta_{y}}}],
\end{equation}\vspace{-1mm}
where $f_n$ is the frequency of the $n$th component, $x_f$ and $t_f$ are the predefined, linear theory estimates of location and time of breaking respectively, and $\theta_y = \arctan(k_{n,y}/k_{n,x})_y$ is the angle of incidence of each wave component at various transverse locations with $\cos \theta_y =0$ in the 2-D and $\cos \theta_y = |y|/\sqrt{x_f^2+y^2}$ in 3-D breaking cases. The discrete frequencies $f_n$ were uniformly spaced over the band $\Delta f = f_N - f_1$ with a central frequency defined by $f_c = {1}/{2}(f_N + f_1)$.  
Following the set up of the initially bimodal wave trains in \citet{Banner-Peirson:2007}, the free surface elevation for a 2-D modulated wave train at the wavemaker is given by\vspace{-1mm}
\begin{equation}
\eta (0,y,t) = a_1 \cos(\omega_1t) + a_2 \cos(\omega_2t -\frac{\pi}{18}), 
\end{equation}
where $\omega_1=2\pi f_1$, $\omega_2=\omega_1+2\pi\Delta f$, and $a_2/a_1 = 0.3$.   Increasing the global steepness $S = \sum _{n=1}^{N} a_nk_n$ increases the strength of the resulting breaking event in both focused packets and modulated wave trains. 
Finally, fluid velocities for each component are calculated using linear theory and then superimposed at the wavemaker. Table \ref{tab1} summarizes the input parameters for all simulated cases.

\section{A new parameterization for the breaking strength parameter $b$}
\begin{figure}
\centering
\includegraphics[width=1\textwidth]{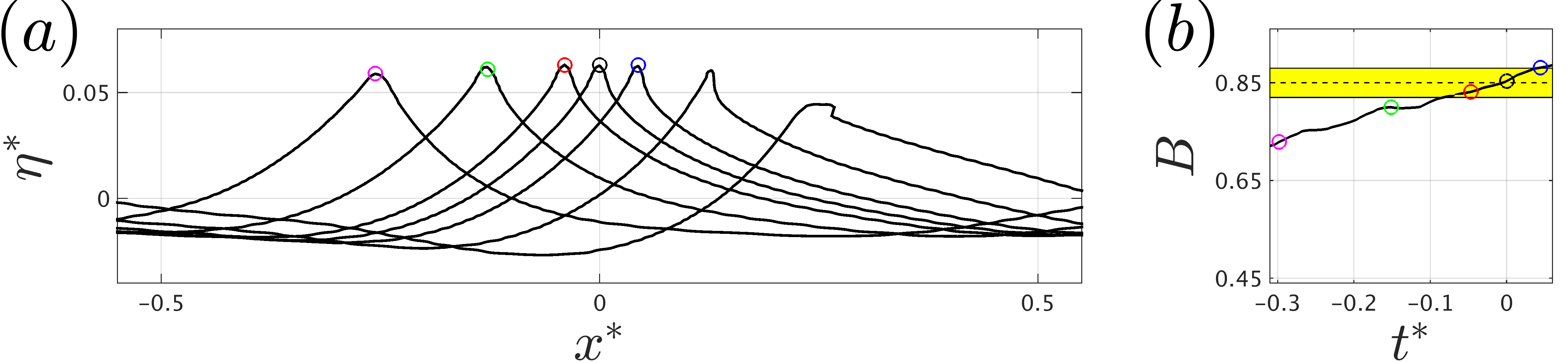}
\includegraphics[width=1\textwidth]{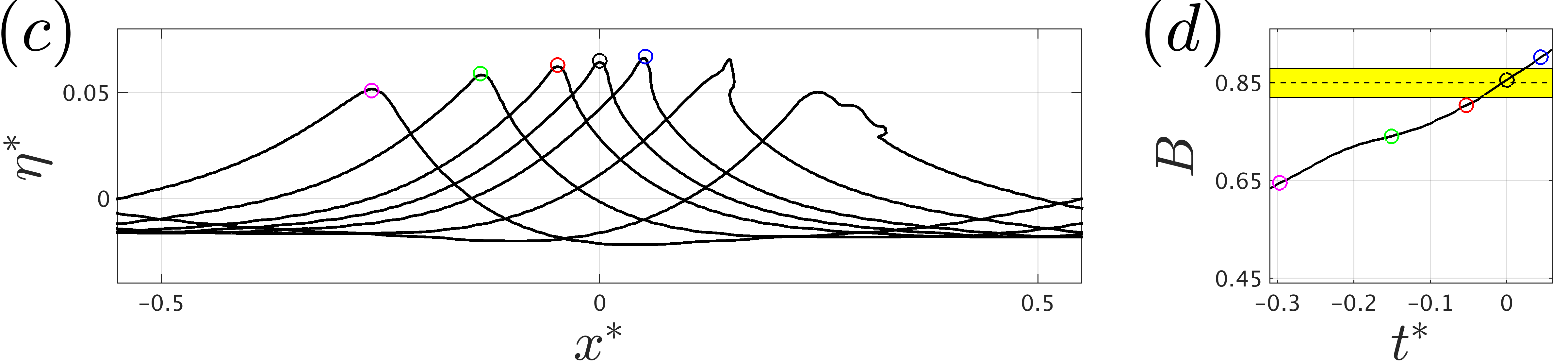}
\includegraphics[width=1\textwidth]{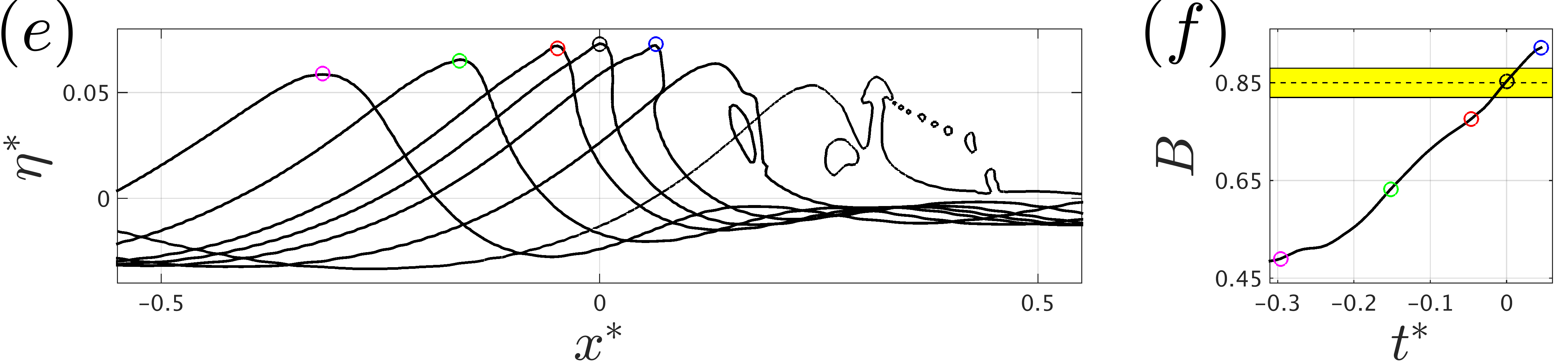}
\caption{$(a,c,e)$ Snapshots of the spatial variation of the normalized free surface elevations near the maximum crest, and $(b,d,f)$ the temporal variation of $B$ before and after the breaking onset for the weak spilling A3, spilling A4 and strong plunging A6 breaking 2-D focused packets.}
\label{fig1}
\end{figure}

\begin{figure}
\centering
\includegraphics[width=1.0\textwidth]{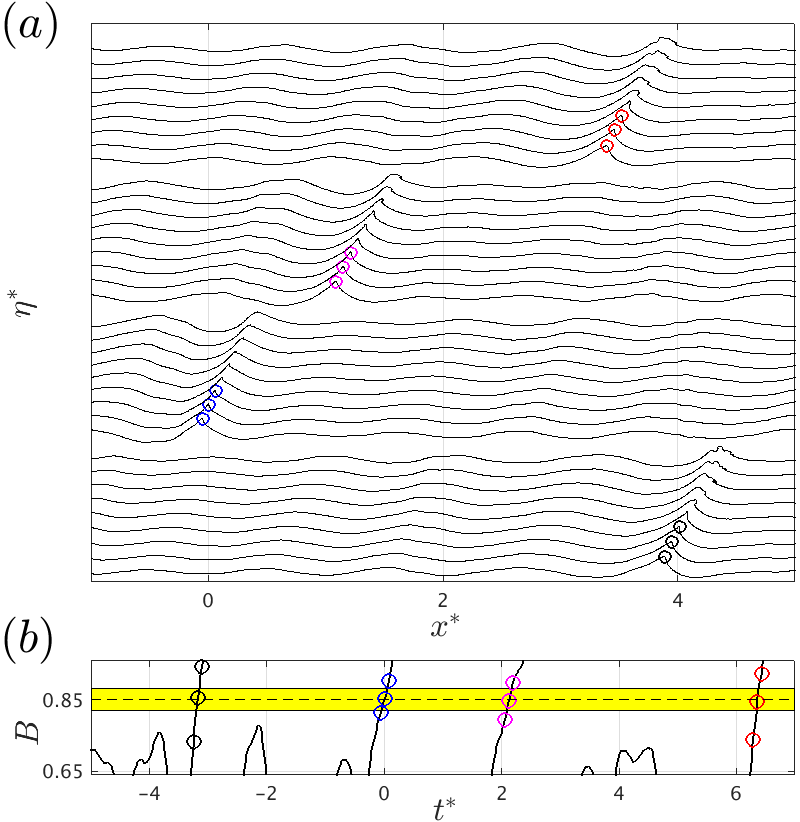}
\caption{$(a)$ Snapshots of the spatial variation of the normalized free surface elevations near the maximum crest, and $(b)$ the temporal variation of $B$ for four successive breaking events in the 2-D multiple-breaking modulated wave train M2.} 
\label{fig2}
\end{figure}

Figure \ref{fig1} shows snapshots of free surface elevations before and after breaking onset as well as the temporal variation of $B$ for the evolving wave crest for the single-breaking focused packets A3, A4 and A6. Figure~\ref{fig2} shows the corresponding results for the multiple-breaking modulated wave train M2. 
Results show that as the strength of breaking increases, the rate of change of $B$ near the threshold value, $dB/dt|_{B_{th}}$, increases. Consistent with \citet{Barthelemy-etal:2018}, warning of imminent breaking onset ($t^*=0$ here) is up to a fifth of a carrier wave period prior to a breaking event. As a consequence, the wave form at $B = B_{th}$ is well defined and the free surface is single-valued. 

\begin{figure}
\centering
\includegraphics[width=\textwidth]{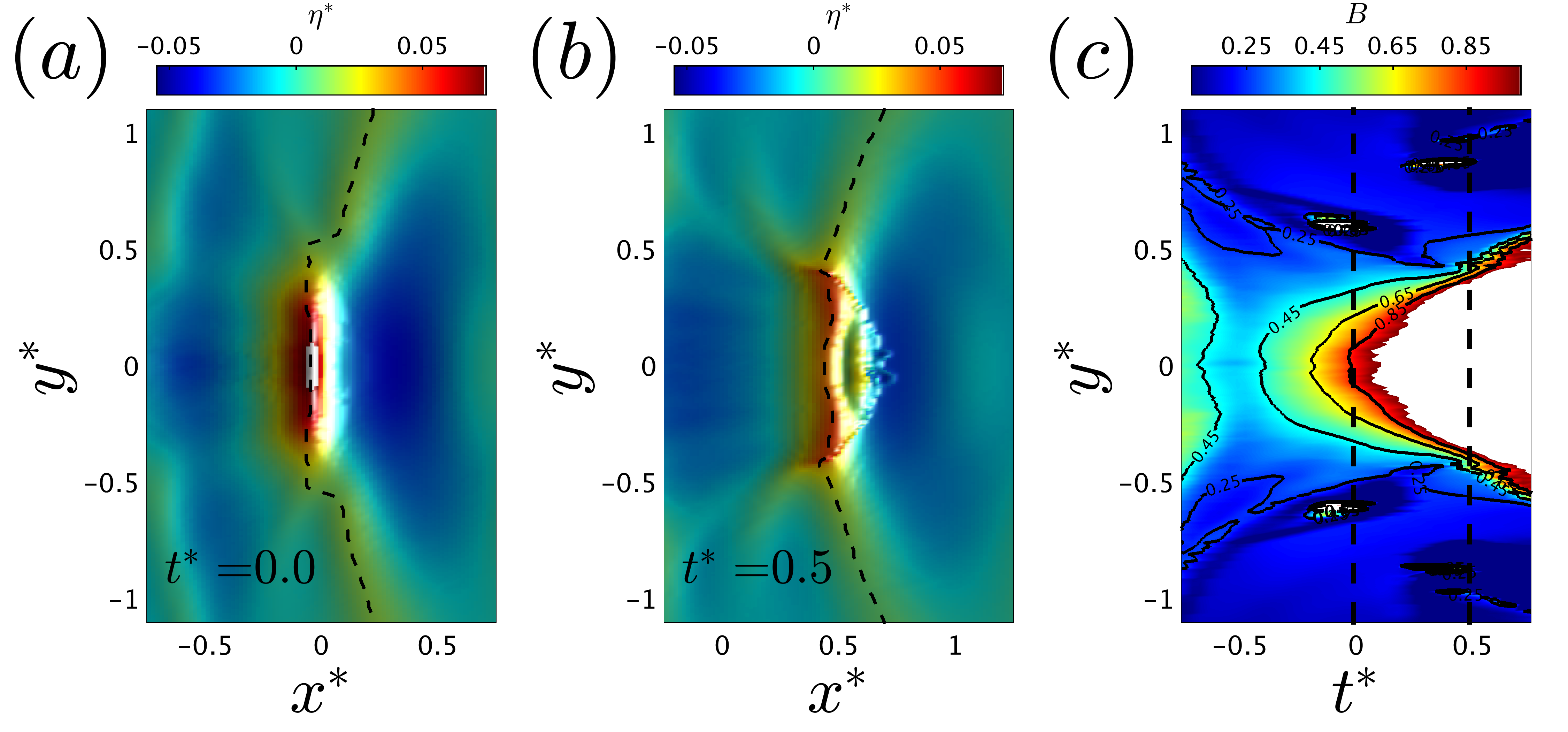}\\
\includegraphics[width=\textwidth]{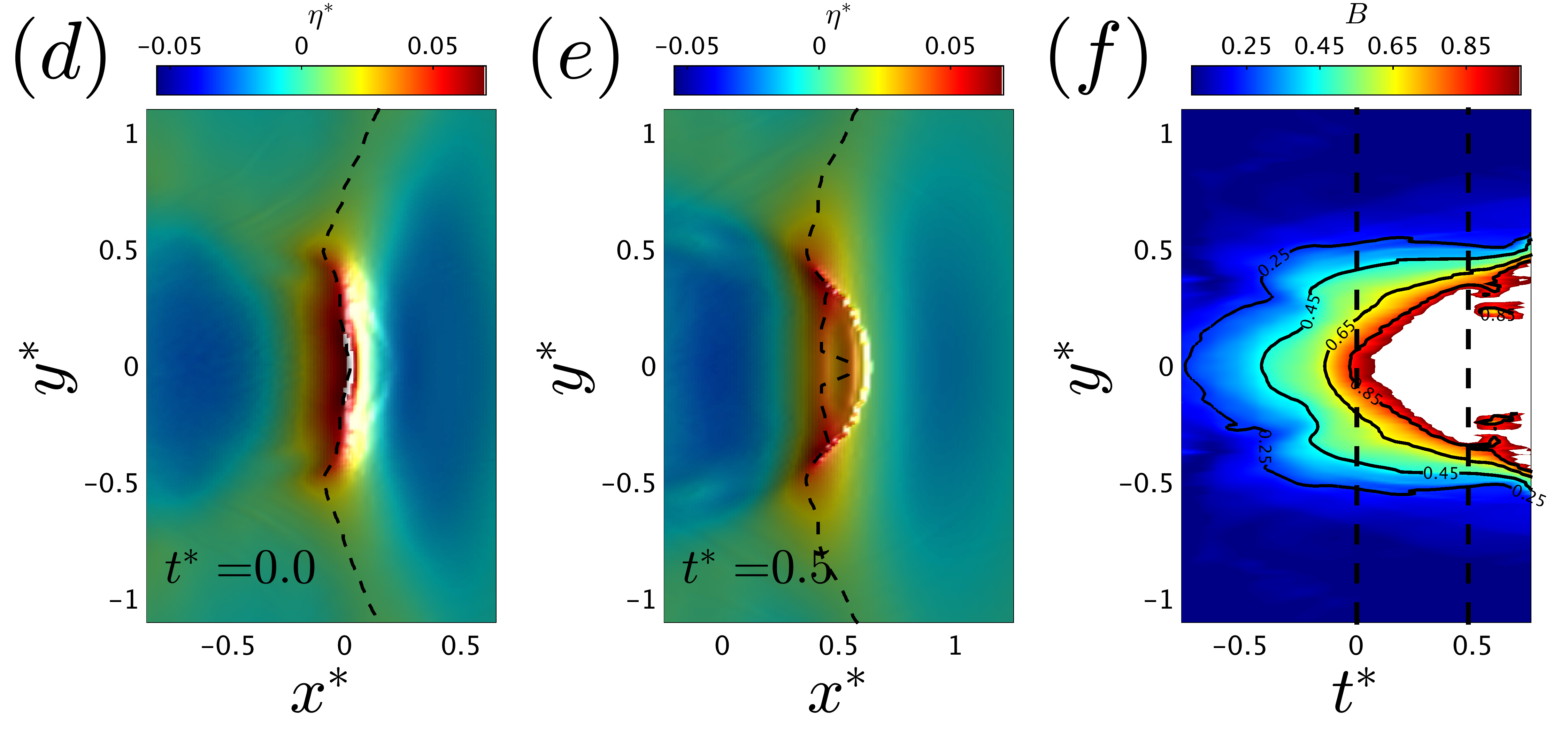}\\
\includegraphics[width=\textwidth]{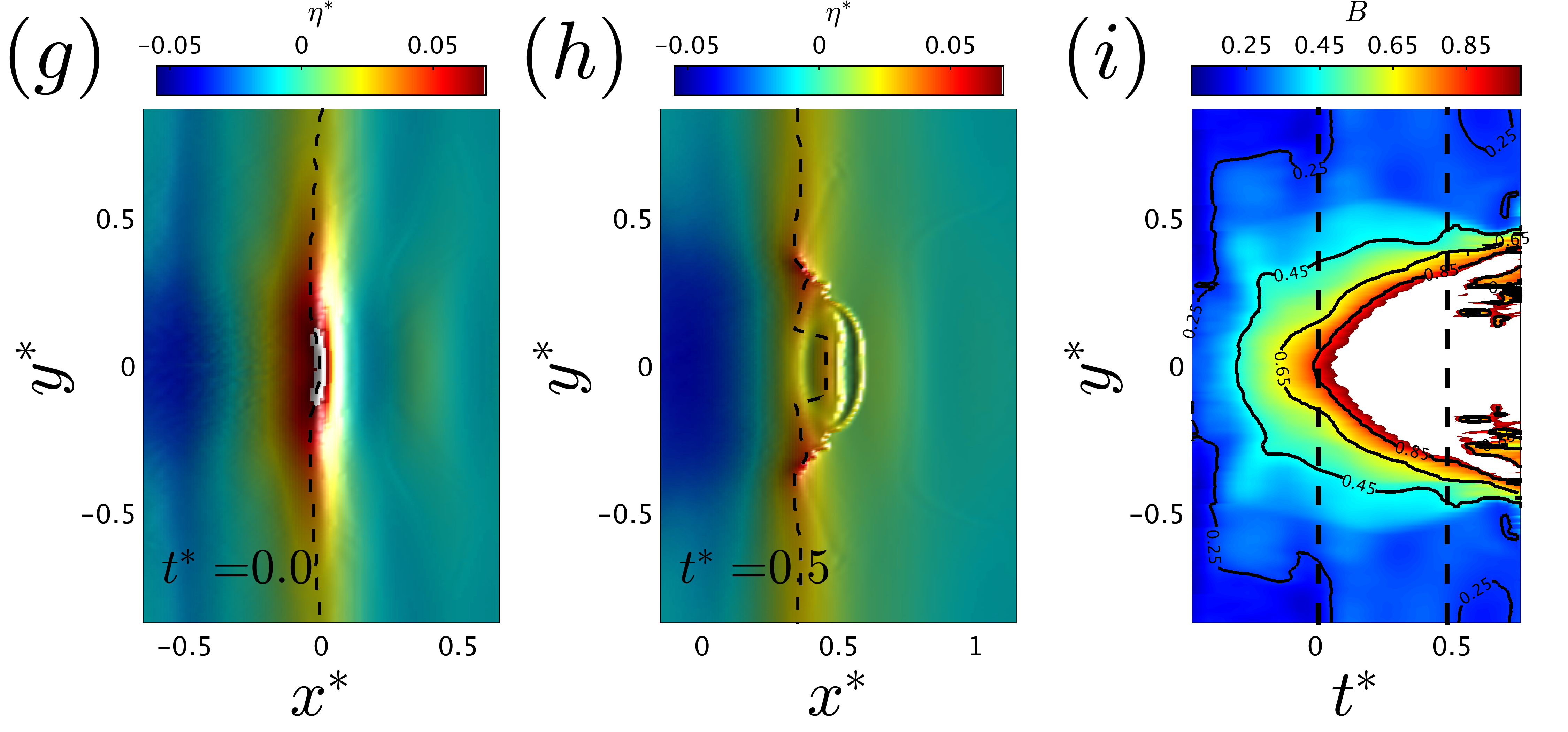}
\caption{Plan view of the spatial variation of the normalized free surface elevations near the maximum crest at $(a,d,g)$ $t^*= 0$, and $(b,e,h)$ $t^* = 0.5$, as well as $(c,f,i)$ the spatiotemporal variation of $B$ for the 3-D focused packets $(a,b,c)$ B1, $(d,e,f)$ B2, and $(g,h,i)$ B4. The corresponding crest locations of the two vertical dashed lines in the right frames are shown by the dashed lines in the left and middle frames in each case. The colors show $0.1<B<1$.}
\label{fig3}
\end{figure}

Figure~\ref{fig3} shows a plan view of the spatial distribution of the normalized free surface elevations at $t^*=0$ at which $B(y\sim 0) = B_{th}$ and after breaking onset at $t^* = 0.5$ as well as the spatiotemporal variation of $B$ for the 3-D breaking focused packets B1, B2 and B4. The latter picture is constructed by displaying the corresponding $B$ curves at each transverse location.  In each case, the dashed lines in the left and middle columns of Figure~\ref{fig3} correspond to the two vertical dashed lines shown in the right column and show the location of the crest maximum at the associated times. 
Due to a strong 3-D focusing of the incident wave packet, the location of the crest maximum at each transverse location may experience rapid change even close to the breaking location, resulting in unrealistic $dB /dt$ values.  Panel $(a)$ demonstrates an example of such a jump in the location of the maximum crest  at $|y^*|\sim 0.6$. The corresponding high values of $B$ shown in the panel $(c)$ for $|y^*|\sim 0.6$ are then just the artifact of the post-processing method.  Figure~\ref{fig3} also shows that the breaking process starts around $y^*=0$, with the breaking crest gradually growing in width during the active breaking period, and that  $B$ has approximately the same temporal structure within the region close to the crest maximum location prior to breaking onset ($|y^*|<0.15$, $-0.1<t^*$). Hereafter we choose $B(y^* = 0)$ as the representative $B$ value for the 3-D cases to compare with the 2-D cases for the proposed parameterization discussed below. 

\begin{figure}
\centering
\includegraphics[width=1.0\textwidth]{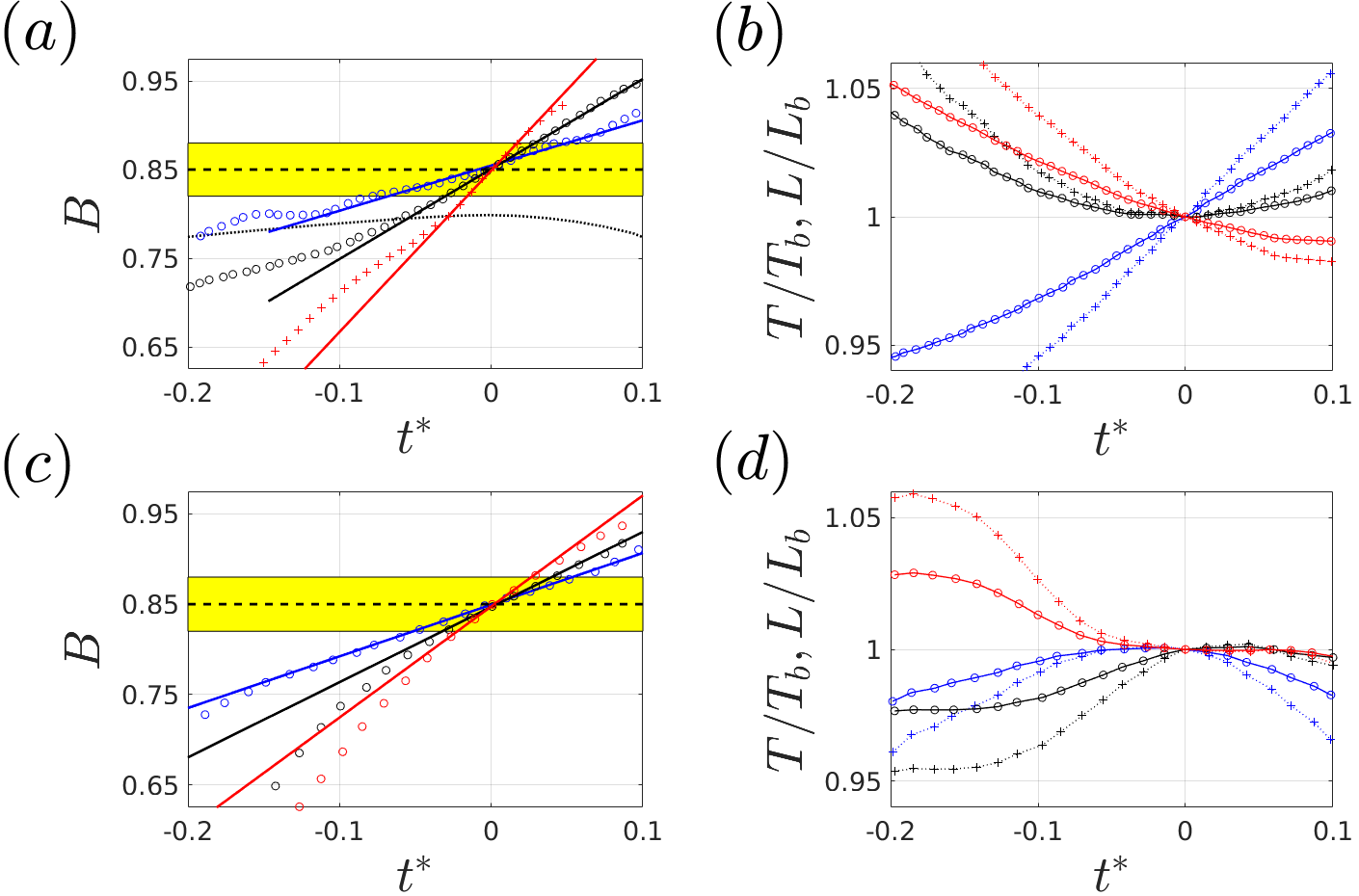}
\caption{Temporal variation of $(a,c)$ $B$, and $(b,d)$ (dashed lines with markers) the local wave length $L$ defined based on the two successive zero crossing point around the crest maximum and (solid lines with markers) local wave period $T$ obtained by using linear dispersion relation and $L$ normalized by their corresponding values at $B = B_{th}$ for $(a,b)$ 2-D focused (dashed-line) non-breaking case A2, and various breaking cases A3, A4, and A6 (markers), as well as $(c,d)$ three successive breaking events in 2-D multiple-breaking modulated wave train M2. In $(a,c)$, solid lines show the linear fit in the interval $|B-B_{th}|<0.03$ shown by the yellow boxes.}
\label{fig4}
\end{figure}

Panels $(a)$ and $(c)$ of Figure~\ref{fig4} present the temporal variation of $B$ for a number of breaking events due to dispersive focusing and modulational instability with various breaking strengths together with a non-breaking case A2, which has a global steepness slightly smaller than the weakest breaking focused packet A3 (see Table~\ref{tab1}). For breaking cases, a linear fit in the interval $|B-B_{th}|<0.03$, shown by the yellow boxes, is also presented. The slope of each such fitted line, hereafter referred to as $dB/dt |_{B_{th}}$, will be used to parameterize the breaking strength parameter $b$. 

To construct a non-dimensional parameter, we also need to identify an appropriate time scale, for which we choose the local period, $T_b$, of the carrier wave at $B = B_{th}$, obtained by using the linear dispersion relation and the local wave length $L_b$ defined based on the two successive zero up- and zero down-crossing points around the crest maximum (see Figure~\ref{fig6}). Panels $(b)$ and $(d)$ of Figure~\ref{fig4} show that such estimation of $L_b$ and $T_b$ have a small variation as the crest approaches breaking onset. 

\begin{figure}
\centering
\includegraphics[width=1.0\textwidth]{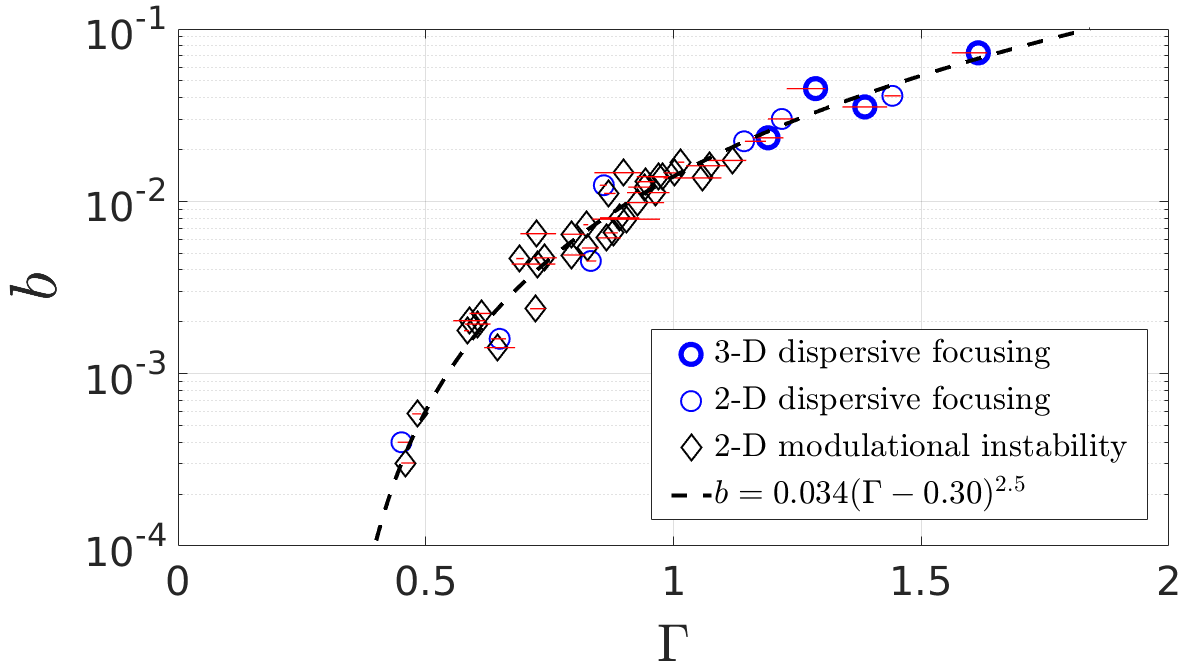}
\caption{Variation in the breaking strength parameter $b$ with the new parameter $\Gamma = T_bdB/dt |_{B_{th}}$ that represents the normalized rate of change of $B$ as it transitions through the breaking onset.}
\label{fig5}
\end{figure}

Figure~\ref{fig5} shows the variation of the breaking strength parameter $b$ with the new parameter $\Gamma = T_bdB/dt |_{B_{th}}$ for all simulated cases. Our methodology to calculate the breaking strength parameter $b$ is similar to that used in \citet{derakhti-kirby-jfm16} with a modification explained in the Appendix A. The reader is referred to \citet{Derakhti-Kirby:2014,derakhti-kirby-jfm16} for the detailed examination of the model prediction of the total dissipation rate compared with the corresponding measured data, as well as the sensitivity of the simulation results with respect to the selected grid resolution. For each case, the horizontal line shows the variation of $\Gamma$ with respect to the selected interval $|B-B_{th}|<\alpha$,  here we set $\alpha = 0.02,0.03,0.04$, and 0.05, to perform the linear fit to obtain $dB/dt |_{B_{th}}$, and the marker represents the associated averaged value of $\Gamma$. Results show that the rate of change of $B$ following the crest tip as it transitions through the breaking onset, and thus the parameter $\Gamma$, are not sensitive to a specific choice of $\alpha$ for $\alpha\le0.05$. Finally, the results show that $\Gamma$ can successfully predict the breaking strength parameter $b$. Assuming a formulation with the form $b \sim a_1(\Gamma-a_2)^{a_3}$ and using the least-square curve fitting technique, we obtain
\begin{equation}
b = 0.034(\Gamma-0.30)^{2.5}. \label{eq3.1}
\end{equation}

\section{Discussion and Conclusions}
The results shown in Figure~\ref{fig5} establish that the strength of individual breaking events in an irregular wave train, as well as the onset of breaking, can be estimated from properties of the wave crest as it approaches breaking.  This is the centerpiece figure of this paper and shows a systematic collapse of the proposed breaking strength predictor $\Gamma$ for a diverse range of representative 2-D water wave focused packets and modulated wave trains in deep and intermediate depth waters, as well as several cases of 3-D dispersive focusing packets.   

The success of the parameterization also would make it possible to better describe breaking events in codes based on potential flow theory, such as high-order spectral (HOS) codes, where breaking is not predicted by the model itself \citep[see][for a recent review of the HOS approach]{ducrozet-etal-oe17}.  The development of criteria for the onset and strength of breaking in such models has long been a subject for investigation.  Recently, \citet{Seiffert-etal:2017} have investigated the use of the parameter $B$ as a breaking onset criterion in HOS, while \citet{Seiffert-Ducrozet:2018} discuss the specification of an eddy viscosity model after the onset of breaking is identified. It is our belief that the specification of such a dissipation model should be based on the parameterization of total dissipation in terms of the rate of change of $B$ developed here, which would provide a strong link between the present work and operational wave modeling.  

Further, the universality of the $B$ parameter and its rate of change as robust indicators of wave breaking onset and strength clearly require further study. For shoaling wave packets, our preliminary results show that the breaking onset conforms to $B\sim0.85$, yet regarding breaking strength, its expression in terms of $b$ is not appropriate. These aspects will be reported in a subsequent paper. We will also address the closely related topic of criteria for breaking of solitary waves shoaling on simple beach geometries for which there is a large body of literature \citep[][for example]{grilli-etal-ww97}. We close by reiterating that the present work provides a first indication of a direct link between the local properties of a wave crest as it transitions through an apparently generic breaking threshold, and the resultant overall energy dissipation resulting from the breaking event. The results are underpinned by fundamental energy flux considerations.

\begin{acknowledgments}
\noindent
{\bf Acknowledgments:} This work was supported by the National Science Foundation, Physical Oceanography Program grant OCE-1435147, and through the use of computational resources provided by Information Technologies at the University of Delaware. M.B. also gratefully acknowledges support from the Australian Research Council for his breaking waves research. The LES/VOF code is based on the model code TRUCHAS, provided by the Los Alamos National Laboratory, Department of Energy.
\end{acknowledgments}

\appendix
\section{Determination of the breaking strength parameter $b$}
\begin{figure}
\centering
\subfigure{\includegraphics[width=0.9\textwidth]{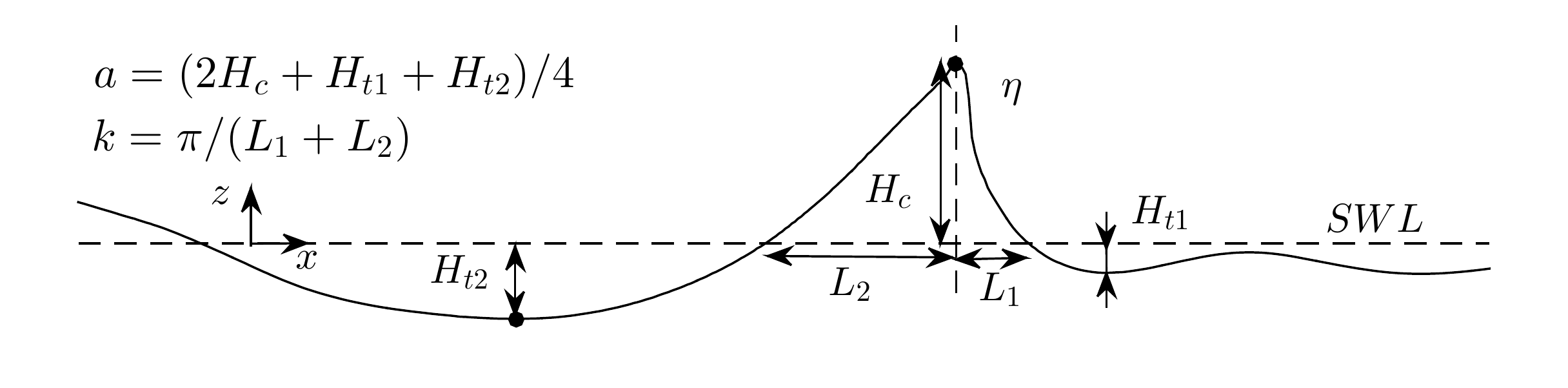}}
\caption{Definitions of the local wave parameters. $k$ and $a$ are the local wave number and amplitude respectively. The wave propagates from the left to the right.}
\label{fig6}
\end{figure}

Rearranging (\ref{eq1.1}), the breaking strength parameter $b$ is written as
\begin{equation}
b = \frac{g\epsilon}{\rho c_b^5}= \frac{g\tilde{\epsilon}}{\rho c_b^5\tau l},   \label{eqA1}
\end{equation} 
where $\tilde{\epsilon}$ is the total wave energy dissipation due to wave breaking, $\tau\sim O(T_b)$ is a time scale related to the active breaking period, and $l$ is mean length of the breaking crest during the active breaking period, $0<t-t_b<\tau$. Following \citet{derakhti-kirby-jfm16} and \citet{tian-etal-pf08}, we estimate the local wavenumber at breaking onset, $k_b$, defined based on the two successive zero up- and zero down-crossing points around the crest maximum as shown in Figure \ref{fig6}. Then, the linear dispersion relation is used to estimate the breaking wave phase speed and period as $c_b = (g/k_b\tanh{k_bh})^{1/2}$ and $T_b=2\pi/k_bc_b$ respectively. 

As described by \citet[][\S4.3]{Derakhti-Kirby:2014}, the dissipation rate during active breaking has a strong spatiotemporal variation, and thus $\epsilon = \tilde{\epsilon}/\tau l$ may be interpreted as an averaged dissipation rate in the interval $0<t-t_b<\tau$. Following \citet{Derakhti-Kirby:2014,derakhti-kirby-jfm16}, we set $\tau = 0.75T_b$.

\citet{derakhti-kirby-jfm16} describe a methodology for computing $\tilde{\epsilon}$ in focused wave packet experiments that depends on the spatial isolation of breaking events and on the fact that the domain is essentially quiescent before and after passage of the wave train. For multiple-breaking cases, and, particularly for the modulational instability cases considered here, the spatial extent of breaking events may overlap in time, rendering the integral-over-all-time approach described by \citet{derakhti-kirby-jfm16} inapplicable. A modification to allow for localization of the total energy loss estimate in both space and time is described here.
Starting from a local equation for mechanical energy $e(x,y,z,t)$ and energy flux ${\bf f} = {\bf u} (p + e)$ per unit volume,
\begin{equation}
e_{,t} + \nabla \cdot {\bf f} = {\epsilon^\prime}  \label{eqB1}
\end{equation}
where ${\epsilon^\prime}$ represents dissipation/unit volume,  we consider the breaking event to be isolated within a region $x_1 \leq x \leq x_2, y_1 \leq y \leq y_2, t_1 \leq t \leq t_2$.  Integrating (\ref{eqB1}) over depth and then over $x$, $y$ and $t$ gives
\begin{eqnarray}
\left[ \int_{x_1}^{x_2} \int_{y_1}^{y_2} \int_{-h}^\eta e dx dy dz \right]^{t_2}_{t_1} & + &
\left[ \int_{y_1}^{y_2} \int_{t_1}^{t_2} \int_{-h}^\eta f_x dy dt dz \right]^{x_2}_{x_1} +
\left[ \int_{x_1}^{x_2} \int_{t_1}^{t_2} \int_{-h}^\eta f_y dx dt dz \right]^{y_2}_{y_1} = \nonumber \\
& = &\int_{x_1}^{x_2} \int_{y_1}^{y_2} \int_{t_1}^{t_2} \int_{-h}^{\eta} {\epsilon^\prime} dx dy dz dt = \tilde{\epsilon}, 
\end{eqnarray}
where $t_1 \le t_b$ and $\tau \ll t_2-t_b$, where $t_b$ is the time of breaking onset as defined in \S2.

\bibliography{Derakhti-etal-jfm-2018}
\bibliographystyle{jfm}

\end{document}